\documentclass[prd,preprint,preprintnumbers,amsmath,amssymb,nofootinbib,nolongbibliography]{revtex4-2}
\usepackage{epsf}
\usepackage{natbib}
\usepackage{graphicx}
\usepackage{dcolumn}
\usepackage{bm}
\usepackage{bbold}

\usepackage{centernot}

\usepackage{amsmath}
\usepackage{slashed} 
\usepackage[font=small,labelfont=bf,figurename=Figure]{caption} 
\usepackage{layout}

\def\be{\begin{equation}}
\def\ee{\end{equation}}

\begin{document}
%
\title[]{Electromagnetic interactions in elastic neutrino-nucleon scattering}
\author{Konstantin A. Kouzakov}
\affiliation{Department of Nuclear Physics and Quantum
Theory of Collisions, Faculty of Physics, Lomonosov Moscow State University, Moscow 119991, Russia}%
\email{kouzakov@srd.sinp.msu.ru}
\author{Fedor M. Lazarev}
\affiliation{Department of Theoretical Physics, Faculty of
Physics, Lomonosov Moscow State University, Moscow 119991, Russia}%
\email{lazarev.fm15@physics.msu.ru}
\author{Alexander I. Studenikin}
\affiliation{Department of Theoretical Physics, Faculty of
Physics, Lomonosov Moscow State University, Moscow 119991, Russia}%
\email{studenik@srd.sinp.msu.ru}

\date{\today}

\begin{abstract}
A thorough account of electromagnetic interactions of massive Dirac neutrinos as well as their spin-flavor state in the theoretical formulation of elastic neutrino-nucleon scattering is given. The formalism of neutrino charge, magnetic, electric, and anapole form factors defined as matrices in the mass basis is employed under the assumption of three-neutrino mixing. The flavor and spin change of neutrinos propagating from the source to the detector is taken into account in the form of a spin-flavor density matrix of the neutrino arriving at the detector. The potential effects of the neutrino charge radii, magnetic moments, and spin polarization in the neutrino-nucleon scattering experiments are outlined.
\end{abstract}

\maketitle



%
\section{Introduction}
\label{intro}
Nonzero masses and mixing of neutrinos indicate that they have properties and interactions beyond the Standard Model (SM), in which neutrinos can couple to $W^\pm$ and $Z^0$ bosons only. In view of this, searches for neutrino electromagnetic (EM) properties are of particular interest~\cite{Window, AnnPhys2016,VMU2020,ARNS2025}. Indeed, it is known that the neutrino charge radii are predicted to be nonzero even within the SM~\cite{Bernabeu00,Bernabeu02,*Bernabeu02Erratum,Bernabeu04}, while nonzero neutrino magnetic moments arise already in the minimally extended SM~\cite{Fujikawa1980}. Also, the beyond-SM theories discuss properties of right-handed neutrinos~\cite{RHN}. In the present study, we consider the most general case of Dirac neutrino EM properties: in addition to the charge radius and magnetic moment, we take into account the electric millicharge and anapole and electric dipole moments.

The effects of neutrino EM properties can manifest themselves both in astrophysical processes, where neutrinos propagate in strong magnetic fields and dense matter, and in laboratory experiments that employ neutrino fluxes from various sources. In the latter case, measurements of cross sections for neutrino scattering on various targets are a rather sensitive and widely used method. For example, Alvarez Ruso \emph{et al}~\cite{vScattering} and Chen~\cite{EFT_for_DM_and_v} discussed neutrino-nucleon scattering at various energies and emphasized that precise theoretical calculations are required to meet the needs of experiments. In order to study experimentally processes of neutrino-nucleon and neutrino-nucleus scattering, it is also necessary to take into account radiative corrections~\cite{Tomalak21JHEP,Tomalak22NatCom,Tomalak22PRD,NCRRadCorCorona24} and details of the inner structure of nucleons and nuclei for both elastic~\cite{Sufian20,Megias20,Zhang20} and inelastic~\cite{Liang20} collisions.

In recent years, great attention has been drawn to coherent elastic neutrino–nucleus scattering (CE$\nu$NS), which was predicted half a century ago~\cite{Freedman74,Kopeliovich74} and which was first experimentally registered in 2017 by the COHERENT collaboration \cite{Akimov}. CE$\nu$NS is not only a new tool for studying neutrino properties, but it also permits probing the structure of nuclei \cite{Yang,Payne,Hoferichter}. Moreover, the CE$\nu$NS process may contribute to the background signal in the experiments searching for dark-matter particles.

Effects of neutrino EM interactions are among expected manifestations of new physics in the CE$\nu$NS experiments. The data from the CE$\nu$NS experiments, such as  COHERENT, CONUS, and Dresden-II, were already used to set limits on the neutrino millicharge, charge radius, and magnetic moment~\cite{NCRRadCorCorona24,Cadeddu2018,*Cadeddu2018Erratum,Miranda,Cadeddu2020,CONUS2022,Dresden-II}. Data from a number of other new experiments aimed at studying CE$\nu$NS are expected to appear in the near future.
In order to study neutrino EM properties in CE$\nu$NS experiments, it is necessary to develop a theoretical formalism that would take into account various neutrino and nuclear EM form factors. It is reasonable to begin solving this problem by considering elastic neutrino scattering on the basic constituent of a nucleus, namely a nucleon. Besides, a positively charged nucleon, i.e., a proton, is the simplest nuclear target. It should be also noted in this connection that elastic neutrino-proton scattering is a promising tool for detecting neutrinos from supernovas~\cite{JUNOYB}. 

The most studied among the neutrino EM properties is the neutrino magnetic moment. One of its marked manifestations is neutrino spin oscillations in a magnetic field. This effect can be significant both in the environments of neutron stars and supernovas and in interstellar and/or intergalactic magnetic fields (see, for instance, Refs.~\cite{PopovEPJC2019,PopovPRD2021} and references therein). Such oscillations affect not only the flavor composition of neutrino fluxes but also the neutrino spin state. This means that a consistent treatment of the neutrino magnetic moment effects in the EM neutrino-nucleon scattering should also account for the neutrino spin polarization due to to the neutrino spin oscillations on the source-detector distance. 

The paper is organized as follows. In Sec.~\ref{nucleon_FF}, we discuss the structure of the vertex function for a spin-1/2 particle, taking into account weak neutral form factors for nucleons and EM form factors for nucleons and neutrinos and present a parametrization of the nucleon form factors, which are used further in numerical calculations. In Sec.~\ref{theory}, we calculate the cross section for elastic neutrino–nucleon scattering under conditions characteristic of neutrino scattering experiments. In Sec.~\ref{results}, we discuss the results of numerical calculations for the neutrino-nucleon cross sections with different initial neutrino spin states on the basis of the formulas obtained for various values of neutrino EM properties and compare them with respective SM predictions. Section~\ref{summary} summarizes this work.

\section{Neutrino and nucleon form factors}
\label{nucleon_FF}
Let us briefly touch upon a general approach to describing the properties of EM and weak interactions of spin-1/2 fermions (for a detailed treatment of this issue in the case of neutrino EM interactions, we refer the interested reader to Ref.~\cite{Window}). The parameters of the fermion-photon and fermion-$Z^0$ boson interactions are specified by the vertex function $\Lambda^{fi}_\mu(p_i,p_f)$, where $p_{i(f)}$ is the momentum of the initial (final) spin-1/2 particle with mass $m_{i(f)}$. This vertex function determines the matrix element of the fermionic current
\begin{equation}
\label{StartingCurrent}
     \langle p_f|j_\mu(0)|p_i\rangle=\bar{u}_{f}(p_f)\Lambda_\mu^{fi}(q,l)u_{i}(p_i),
\end{equation}
where $u_{i}(p_i)$ and $u_{f}(p_f)$ are bispinor amplitudes for a free particle.
Employing $q^\mu=p^\mu_i-p^\mu_f$, $l^\mu=p^\mu_i+p^\mu_f$, the metric tensor $g^{\mu\nu},$ and the Levi-Cevita antisymmetric tensor $\varepsilon^{\mu\nu\alpha\beta}$, the $4\times4$ matrix $\Lambda_\mu^{fi}(q,l)$ can be presented as a linear combination of the $16$ matrices $\mathbb{1},$ $\gamma_5,$ $\gamma^\mu,$ $\gamma^\mu\gamma_5$ and $\sigma^{\mu\nu}=\frac{i}{2}[\gamma^\mu,\gamma^\nu]$. From hermiticity of the current operator $j^\dagger_\mu=j_\mu$ it follows that
\begin{equation}
\label{Hermiticity}
     [\Lambda_{\mu}^{fi}(q,l)]^\dagger=\gamma^0\Lambda_\mu^{if}(-q,l)\gamma^0.
\end{equation}
 After using the Gordon-like identities one arrives at~\cite{Nowakowski}
\begin{equation}
\label{GeneralNCFF}
\begin{aligned}
    \Lambda^{fi}_\mu(q)=&f^{fi}_1(q^2)q_\mu+f^{fi}_2(q^2)q_\mu\gamma_5+f^{fi}_3(q^2)\gamma_\mu+f^{fi}_4(q^2)\gamma_\mu\gamma_5+f^{fi}_5(q^2)\sigma_{\mu\nu}q^\nu\\
&+f^{fi}_6(q^2)\varepsilon_{\mu\nu\alpha\beta}q^\nu\sigma^{\alpha\beta}.
\end{aligned}
\end{equation}
By virtue of the requirement of Lorentz invariance, the six form factors $f^{fi}_{1,2,3,4,5,6}(q^2)$ appearing here are functions of $q^2$. Of them, $f^{fi}_{2,3,4}(q^2)$ are hermitian matrices, while $f^{fi}_{1,5,6}(q^2)$ are antihermitian.

In the case of the EM current, the gauge invariance implies the current conservation $\partial_\mu j^\mu=0$, so that
\begin{equation}
\label{CurrentConservation}
     q^\mu\bar{u}_{f}(p_f)\Lambda_\mu^{fi}(q,l)u_{i}(p_i)=0.
\end{equation}
This condition (along with the Gordon-like identities) leads to
\begin{equation}
\label{CurrentConservationFF}
\begin{aligned}
      f^{fi}_1q^2+f^{fi}_3(m_f-m_i)=0, \qquad
      f^{fi}_2q^2+f^{fi}_4(m_f+m_i)=0.
\end{aligned}
\end{equation}
Therefore, the EM vertex acquires the form
\begin{equation}
\label{EMFFs}
     \Lambda^{({\rm EM})fi}_\mu(q)=\left(\gamma_\mu-\frac{q_\mu\slashed{q}}{q^2}\right)\left[f^{fi}_3(q^2)-\frac{q^2}{m_i+m_f}\,f^{fi}_2(q^2)\gamma_5\right]+\sigma_{\mu\nu}q^\nu[f^{fi}_5(q^2)-2if^{fi}_6(q^2)\gamma_5].
\end{equation}
\subsection{Neutrino EM form factors}
From expression (\ref{EMFFs}) it follows that the neutrino EM vertex function can be recast as (see also Ref.~\cite{Window})
\begin{equation}
\label{EMvFF}
     \Lambda^{({\rm EM};\nu)fi}_\mu(q)=(\gamma_\mu-q_\mu\slashed{q}/q^2)[f^{fi}_Q(q^2)+f^{fi}_A(q^2)q^2\gamma_5]-i\sigma_{\mu\nu}q^\nu[f^{fi}_M(q^2)+if^{fi}_E(q^2)\gamma_5],
\end{equation}
where $f^{fi}_Q, f^{fi}_A, f^{fi}_M$ and $f^{fi}_E$ are, respectively, the charge, anapole, magnetic, and electric form factors of the diagonal ($f=i$) and transition ($f\neq i$) types in the basis of neutrino mass eigenstates. According to the hermiticity condition~(\ref{Hermiticity}), we have
\begin{equation}
\label{vHermiticity}
     f^{fi}_{Q,A,M,E}(q^2)=(f^{fi}_{Q,A,M,E}(q^2))^*.
\end{equation}
In the interaction with a real photon ($q^2=0$), the EM form factors determine stationary properties of neutrinos:
\begin{equation}
\label{EMvM}
f^{fi}_Q(0)=e_{fi},\quad 6\left.\frac{df^{fi}_Q(q^2)}{dq^2}\right|_{q^2=0}=\langle r^2_{fi}\rangle, \quad f^{fi}_A(0)=a_{fi}, \quad f^{fi}_M(0)=\frac{\mu_{fi}}{2m_e}, \quad f^{fi}_E(0)=\frac{\epsilon_{fi}}{2m_e},
\end{equation}
where $e_{fi}, \langle r^2_{fi}\rangle, a_{fi}, \mu_{fi}$ and $\epsilon_{fi}$ are, respectively, the neutrino millicharge (in the units of elementary charge $e_0$), charge radius\footnote{This term is commonly used in the literature for the mean-square charge radius.} and anapole moment (both are usually measured in cm$^2$), magnetic and electric dipole moments (in the units of Bohr magneton $\mu_B$).

The neutrino EM form factors play a fundamental role. In particular, the neutrino EM properties may shed light on whether neutrinos are Dirac or Majorana particles: for example, Dirac neutrinos may have both diagonal and off-diagonal charge, magnetic, and electric form factors, while Majorana neutrinos may have only off-diagonal form factors of these types.

Among neutrino EM properties, the magnetic and electric dipole moments have been most studied theoretically. For the diagonal magnetic and electric dipole moments of the Dirac neutrino, the minimally extended SM, which takes into account massive right-handed neutrinos, predicts the values~\cite{Fujikawa1980}
\begin{equation}
\label{Minimally Extended SM_MM}
         \mu_{kk}^D\approx3.2\times10^{-19}\mu_{\textrm{B}}\left(\frac{m_k}{1\ \rm \text{eV}}\right), \qquad \epsilon_{kk}^D=0,
\end{equation}
where $\mu_{\textrm{B}}$ is the Bohr magneton. Because of proportionality to the neutrino masses, the magnetic moments are many orders of magnitude smaller than the current experimental limits: $\mu_\nu\lesssim10^{-12}-10^{-11}\,\mu_{\textrm{B}}$ (see \cite{PDG24}). Nevertheless, these EM properties are a subject of searches in various experiments, since some theories beyond the minimally extended SM predict much greater values for them.
%
Also, we would like to indicate the mechanism underlying the appearance of the induced magnetic moment of a massive neutrino that moves in a dense degenerate electron gas. The effective value of this magnetic moment may be several orders of magnitude greater than the value in Eq.~(\ref{Minimally Extended SM_MM}) (see Refs.~\cite{TernovJETP,TernovPRD} for details).

The neutrino charge and anapole form factors are also of interest. The possible existence of a nonzero neutrino electric charge (millicharge) is considered in some theories beyond the SM. Even if the millicharge is zero, the neutrino may have a nonzero charge radius, which contributes to neutrino scattering. Also, the neutrino may have an anapole moment, whose effect on neutrino scattering by a target is similar to the effect of the neutrino charge radius (see Ref.~\cite{Window} for a detailed discussion of this issue).

In the SM with only one generation, the electric neutrality of the neutrino, along with charge quantization of leptons and quarks, is due to the cancellation of gauge anomalies in the case of electroweak interaction. 
However, there is an exotic possibility in the three-generation SM to cancel the gauge anomalies in such a way that two neutrinos have opposite electric charges and the third neutrino is electrically neutral, with the charges of the leptons of two generations being dequantized and the quark charges obeying the standard charge quantization~\cite{Foot1990,Foot1993,SMvMCh}. In the case of theories featuring massive right-handed Dirac neutrinos, which are singlets of the $SU(2)$ weak-isospin group and which have a nonzero hypercharge, scenarios in which the neutrino has a nonzero electric charge are also possible \cite{Babu1990}. As a result, cancellation of gauge anomalies requires the respective shift of charges of charged leptons and quarks and, hence, of electrons, protons, and neutrons. It follows that the neutrino charge should be sufficiently small for the electric neutrality of matter to remain unaltered. This yields the constraint~\cite{NeutOfMatter}
\begin{equation}
\label{Neutrality of matter}
         |e_{ee}|\lesssim3\times10^{-21}\,e_0.
\end{equation}
Even if the electric charge of the neutrino is zero, the charge form factor $f^{fi}_Q(q^2)$ may carry nontrivial information about neutrino electric properties---namely, about the neutrino charge radius. The calculation of the neutrino charge radius within the SM yields diagonal in the flavor basis results\footnote{Our definition of the SM neutrino charge radii differs in sign from that in Refs.~\cite{Bernabeu00,Bernabeu02,Bernabeu04} for the reasons explained in Ref.~\cite{Cadeddu2018}.}~\cite{Bernabeu00,Bernabeu02,Bernabeu04}
\begin{equation}
\label{SM CR formula}
       \langle r^2_{\ell\ell}\rangle_{SM}=-\frac{G_F}{4\sqrt{2}\pi^2}\left[3-2\log{\frac{m_\ell^2}{m_W^2}}\right],
\end{equation}
where $m_W$ and $m_\ell$ are the $W$-boson and charged lepton $(\ell=e,\mu,\tau)$ masses, respectively. From Eq. (\ref{SM CR formula}), one can obtain the following numerical estimates:
\begin{equation}
\label{SM CR values}
    \begin{aligned}
    \langle r_{ee}^2\rangle_{SM}&=-4.1\times10^{-33}\,\textrm{cm}^2,\\
    \langle r_{\mu\mu}^2\rangle_{SM}&=-2.4\times10^{-33}\,\textrm{cm}^2,\\
    \langle r_{\tau\tau}^2\rangle_{SM}&=-1.5\times10^{-33}\,\textrm{cm}^2.
    \end{aligned}
\end{equation}
These values are only one order of magnitude less than the current experimental limits: $|\langle r^2_{\ell\ell}\rangle|\lesssim 10^{-32}\,\textrm{cm}^2$~\cite{PDG24}. It is pointed out in Ref.~\cite{Bernabeu02} that the SM also predicts nonzero values for the neutrino diagonal in the flavor basis anapole moments. According to the vertex decomposition~(\ref{EMvFF}) it is simply 
\begin{equation}
\label{SM vAnapole}
    a^{SM}_{\ell\ell}=-\frac{1}{6}\langle r^2_{\ell\ell}\rangle_{SM}.
\end{equation}
As discussed in Ref.~\cite{Window}, in the SM with massless left-handed neutrinos the neutrino charge radius and the anapole moment are not defined separately, so that one can interpret their linear combination $\langle r^2_{\ell\ell}\rangle_{SM}/6-a^{SM}_{\ell\ell}$, which appears in Eq.~(\ref{EMvFF}) at $q^2\to0$ after transforming to the flavor basis, as either a charge radius or an anapole moment. In this work, we intentionally employ the separate definitions~(\ref{SM CR formula}) and~(\ref{SM vAnapole}), since we use the corresponding values in the calculations that illustrate the combined effects of the neutrino EM properties and spin polarization (see Sec.~\ref{results}).    

\subsection{Nucleon EM and weak neutral form factors}
In describing neutrino-nucleon scattering, we use the nucleon EM vertex function in the standard form~\cite{Nowakowski,Haxton01} [setting $m_f=m_i=m_N$ in Eq.~(\ref{EMFFs}), where $m_N$ is the nucleon mass]
\begin{equation}
\label{EMNFF}
     \Lambda^{({\rm EM};N)}_\mu(q)=\gamma_\mu F_Q^N(q^2)-\frac{i}{2m_N}\,\sigma_{\mu\nu}q^\nu F_M^N(q^2)+\frac{1}{2m_N}\,\sigma_{\mu\nu}q^\nu\gamma_5 F_E^N(q^2)-\left(q^2\gamma_\mu-q_\mu\slashed{q}\right)\gamma_5 \frac{F_A^N(q^2)}{m_N^2}.
\end{equation}
Here $F_{Q,M,E,A}$ are, respectively, charge, magnetic, electric, and anapole form factors for the proton ($N=p$) and the neutron ($N=n$).

In the case of the weak neutral current, we disregard second-class currents, which violate isotopic invariance of strong interaction--that is, we set $f_1=f_6=0$ in Eq.~(\ref{GeneralNCFF}). As a result, we obtain~\cite{Giunti_book,Alberico09}
\begin{equation}
\label{NCNFF}
     \Lambda^{({\rm NC};N)}_\mu(q)=\gamma_\mu F_1^N(q^2)-\frac{i}{2m_N}\,\sigma_{\mu\nu}q^\nu F_2^N(q^2)-\gamma_\mu\gamma_5 G_A^N(q^2)+\frac{1}{m_N}\,G_P^N(q^2)q^\mu\gamma_5,
\end{equation}
where $F_1^N$, $F_2^N$, $G_A^N$, and $G_P^N$ are referred to as, respectively, the nucleon Dirac, Pauli, axial, and pseudoscalar weak neutral form factors. Below, we omit the pseudoscalar form factor $G_P^N$, since its contribution to the neutrino-nucleon cross section vanishes in the limit of zero neutrino mass.

Restricting ourselves, in the description of the internal structure of nucleons, to the $u$, $d$ and $s$ light quarks, for which $SU(3)$ flavor symmetry holds approximately, and employing the hypotheses of vector current conservation and partial axial-current conservation, we can relate nucleon weak neutral currents to nucleon EM currents. This entails relations between the respective form factors~\cite{Giunti_book,Kosmas16} (here, we restrict ourselves to the charge and magnetic form factors in the case of the EM current). Specifically, we have
\begin{equation}
\label{NFFinterconnection}
    \begin{aligned}
    F_{1,2}^p(q^2)&=\left(\frac{1}{2}-2\sin^2\theta_W\right)F_{Q,M}^p-\frac{3}{2}F_{Q,M}^n-\frac{1}{2}F_{1,2}^S,\\
    F_{1,2}^n(q^2)&=\left(\frac{1}{2}-2\sin^2\theta_W\right)F_{Q,M}^n-\frac{3}{2}F_{Q,M}^p-\frac{1}{2}F_{1,2}^S,\\
    G_{A}^p(q^2)&=\frac{1}{2}G_{A}(q^2)-\frac{1}{2}G_{A}^S(q^2),\\
    G_{A}^n(q^2)&=-\frac{1}{2}G_{A}(q^2)-\frac{1}{2}G_{A}^S(q^2),
    \end{aligned}
\end{equation}
where $G_A$ is the axial form factor, while $F_{1,2}^S$ and $G_{A}^S$ are the strange form factors for the nucleon.
\subsection{Parametrization of nucleon form factors}
\label{Parametrization of nucleon form factors}
In the literature, the $q^2$ dependence of the nucleon form factors is frequently described in the dipole approximation. However, the parametrization of the nucleon form factors in the dipole approximation is not always sufficient for performing a detailed analysis of experimental data on lepton-nucleon scattering. For this reason, we make use here of the more accurate approach developed in Refs.~\cite{Kosmas16,Alberico09}.

We introduce the Sachs EM form factors $G^N_{E,M}$,
\begin{equation}
    \begin{aligned}
    F_Q^N(q^2)&=\frac{G_E^N(q^2)-\frac{q^2}{4m_N^2}G_M^N(q^2)}{1-\frac{q^2}{4m_N^2}},
   \qquad F_M^N(q^2)&=\frac{G_M^N(q^2)-G_E^N(q^2)}{1-\frac{q^2}{4m_N^2}},
    \end{aligned}
\end{equation}
which are parametrized as
\begin{equation}
    \begin{aligned}
    \frac{G_M^N(q^2)}{\mu_N}&=\frac{1-\frac{q^2}{4m_N^2}a_M^N}{1-\frac{q^2}{4m_N^2}b_{M1}^N+\left(\frac{q^2}{4m_N^2}\right)^2b_{M2}^N-\left(\frac{q^2}{4m_N^2}\right)^3b_{M3}^N},\\
    G_E^p(q^2)&=\frac{1-\frac{q^2}{4m_N^2}a_E^p}{1-\frac{q^2}{4m_N^2}b_{E1}^p+\left(\frac{q^2}{4m_N^2}\right)^2b_{E2}^p-\left(\frac{q^2}{4m_N^2}\right)^3b_{E3}^p},\\
    G_E^n(q^2)&=\frac{-\frac{q^2}{4m_N^2}\lambda_1}{1-\frac{q^2}{4m_N^2}\lambda_2}\left(1-\frac{q^2}{M_V^2}\right)^{-2},
    \end{aligned}
\end{equation}
where $\mu_N$ is the nucleon magnetic moment in nuclear magneton units. For the axial form factor, we use the parametrization
\begin{equation}
    G_A(q^2)=g_A\left(1-\frac{q^2}{M_A^2}\right)^{-2}.
\end{equation}
We now present the numerical values of all of the
parameters used here~\cite{Alberico09}:
\begin{equation}
    \begin{aligned}
    m_N=938\,\textrm{MeV},&\qquad\mu_p=2.793,\qquad\mu_n=-1.913,\\
    M_V=843\,\textrm{MeV},&\qquad g_A=1.267,\qquad M_A=1049\,\textrm{MeV},\\
    a_E^p=-0.19,\qquad b_{E1}^p&=11.12,\qquad b_{E2}^p=15.16,\qquad b_{E3}^p=21.25,\\
    a_M^p=1.09,\qquad b_{M1}^p&=12.31,\qquad b_{M2}^p=25.57,\qquad b_{M3}^p=30.61,\\
    \lambda_1&=1.68,\qquad\lambda_2=3.63,\\
    a_M^n=8.28,\qquad b_{M1}^n&=21.3,\qquad b_{M2}^n=77,\qquad b_{M3}^n=238.
    \end{aligned}
\end{equation}
The strange form factors are parameterized as~\cite{Garvey93}
\begin{equation}
    \begin{aligned}
    F_1^S(q^2)&=\frac{\frac{q^2}{6}\langle r_S^2\rangle}{\left(1-\frac{q^2}{4m_N^2}\right)}\left(1-\frac{q^2}{M_V^2}\right)^{-2},\\
    F_2^S(q^2)&=\frac{\mu_S}{\left(1-\frac{q^2}{4m_N^2}\right)}\left(1-\frac{q^2}{M_V^2}\right)^{-2},\\
    F_A^S(q^2)&=g_A^S\left(1-\frac{q^2}{M_A^2}\right)^{-2},
    \end{aligned}
\end{equation}
where $\langle r_S^2\rangle$ is the strange nucleon radius, $\mu_S$ is the strange nucleon magnetic moment, and $g_A^S$ is the strange contribution to the nucleon spin. Measurements of strange vector form factors $F^S_{1,2}$ in parity-violating electron scattering indicate their values close to zero \cite{Liu07}. Therefore, in this work, following Ref.~\cite{Butkevich23}, we neglect the contributions of strangeness, i.e., it is assumed that $F^S_{1,2}=0$. However, we take into account the strange contribution to the weak axial form factor. We consider $0.2\geq g^S_A\geq -0.2$, which almost completely covers the experimental values of $g^S_A$ (see, for example, Ref.~\cite{Butkevich23}). In the present study, we will compare the scattering cross sections calculated with and without allowance for the above strange form factors.

\section{Cross sections for elastic neutrino-nucleon scattering}
\label{theory}
We consider the process where an ultrarelativistic neutrino with 4-momentum $k^\mu=(E_\nu,\bm{k})$ originates from a source (reactor, accelerator, the Sun, etc.) and elastically scatters on a nucleon in a detector at energy-momentum transfer $q=(T,\bm{q})$. If the neutrino has nonzero magnetic moment, its spin-flavor oscillations may take place due to interactions with a magnetic field in the source and with interstellar and/or intergalactic magnetic fields. In the most general case, the neutrino state in the detector can be mixed both in flavor/mass space and in spin space. We can describe the neutrino state before scattering by a statistical operator acting in the Hilbert space of neutrino mass and spin states as
\begin{equation}
\label{Dirac density matrix}
\rho_{ij}=\alpha_{ij}u^{(\nu_i)}_{k,-}\bar{u}^{(\nu_j)}_{k,-}+\kappa_{ij}u^{(\nu_i)}_{k,-}\bar{u}^{(\nu_j)}_{k,+}+\kappa_{ji}^*u^{(\nu_i)}_{k,+}\bar{u}^{(\nu_j)}_{k,-}+\beta_{ij}u^{(\nu_i)}_{k,+}\bar{u}^{(\nu_j)}_{k,+},
\end{equation}
where $\alpha,\beta$ are hermitian matrices, $u^{(\nu_i)}_{k,\pm}$ is the bispinor amplitude of the massive neutrino state $|\nu_i\rangle$ with 4-momentum $k$ and positive ($+$) or negative ($-$) helicity, and in the Weyl representation it has the form
\begin{equation}
\label{Dirac bispinors}
u^{(\nu_i)}_{k,\pm}=\begin{pmatrix}\sqrt{(k\cdot\sigma)}\xi_\pm\\\sqrt{(k\cdot\bar{\sigma})}\xi_\pm\end{pmatrix}=\begin{pmatrix}\sqrt{E_\nu(m_i)\mp|\bm{k}|}\xi_\pm\\\sqrt{E_\nu(m_i)\pm|\bm{k}|}\xi_\pm\end{pmatrix}.
\end{equation}
Here $\xi_\pm$ are helisity 2-component eigenspinors, $(\bm{k}\cdot\bm{\sigma})\xi_\pm=\pm|\bm{k}|\xi_\pm$. In the ultrarelativistic limit these bispinor amplitudes are also the chirality eigenstates
\begin{equation}
\label{ultrarelativistic dirac bispinors}
    u^{(\nu_i)}_{k,+}=\sqrt{2E_\nu}\begin{pmatrix}0\\\xi_+\end{pmatrix},\qquad u^{(\nu_j)}_{k,-}=\sqrt{2E_\nu}\begin{pmatrix}\xi_-\\0\end{pmatrix},
\end{equation}
and $E_\nu=|\bm{k}|$. Substituting the explicit form~(\ref{ultrarelativistic dirac bispinors}) in Eq.~(\ref{Dirac density matrix}), we can rewrite the density matrix as a generalization of the ultrarelativistic spin density matrix for a Dirac particle:
\begin{equation}
\label{Landau density matrix}
    \rho_{ij}=\frac{1}{2}\slashed{k}\left(\tilde\rho_{ij}-\zeta^\parallel_{ij}\gamma_5+(\bm{\zeta}^\perp_{ij}\cdot\bm{\gamma}_\perp)\gamma_5\right),
\end{equation}
where $\tilde\rho_{ij}=\frac{1}{2E_\nu}tr(\rho_{ij}\gamma^0)=\alpha_{ij}+\beta_{ij}$ is a reduced density matrix in the neutrino mass space. $\frac{1}{2}\zeta^\parallel_{ij}=\frac{1}{2}(\beta_{ij}-\alpha_{ij})$ and, with $\{{\bm e}_x,{\bm e}_y,{\bm k}/E_\nu\}$ forming a 3-vector basic, ${\bm\zeta}^\perp_{ij}=\zeta^x_{ij}\bm{e}_x+\zeta^y_{ij}\bm{e}_y=(\kappa_{ij}+\kappa^*_{ji})\bm{e}_x-i(\kappa_{ij}-\kappa^*_{ji})\bm{e}_y$ are the matrices of longitudunal ($\parallel$) and transverse ($\perp$) components, with respect to the neutrino momentum $k$, corresponding to the diagonal ($i=j$) and transition ($i\neq j$) values of the spin operator $\bm{\Sigma}$ averaged over spin states of the neutrino in its rest frame. If there is a state with zero neutrino mass then the diagonal and transition ${\zeta}^\perp$ components involving this state vanish. 
Thus, the density matrix (\ref{Landau density matrix}) contains the information 
about all the neutrino spin components, so that, for example, $\frac{1}{2E_\nu}tr(\rho_{ij}\gamma^0\frac{1}{2}\bm{\Sigma)}=\frac{1}{2}\zeta^\parallel_{ij}\frac{\bm{k}}{E_\nu}$ are the values of the spin operator averaged over neutrino spin states in the lab frame (there is only longitudinal part, as it should be for the ultrarelativistic limit). Also we note that the density matrix~(\ref{Dirac density matrix}) can be determined taking into account that in the ultrarelativistic limit the components of the spin-flavor density matrix in the mass basis typically employed in neutrino oscillation calculations are related to the components $\frac{1}{4E^2_\nu}(u^{(\nu_i)}_{k,r'})^\dagger\rho_{ij}\gamma^0u^{(\nu_j)}_{k,r}$. Such a density matrix acting on the column vectors in the basis of neutrino states $\{(\nu^R_1,\nu^R_2,\nu^R_3),(\nu^L_1,\nu^L_2,\nu^L_3)\}$ can be presented as
\begin{equation}
\label{Oscill density matrix}
    \rho=\begin{pmatrix}
    \beta & \kappa^\dagger\\
    \kappa& \alpha
    \end{pmatrix}=\frac{1}{2}\begin{pmatrix}
    \tilde\rho+\zeta^\parallel & \zeta^x-i\zeta^y\\
    \zeta^x+i\zeta^y& \tilde\rho+\zeta^\parallel
    \end{pmatrix}=\frac{1}{2}\bigg(\tilde\rho\mathbb{1}+(\bm{\zeta}\cdot\hat{\bm{\sigma}})\bigg),
\end{equation}
where $\mathbb{1}$ and the Pauli matrices $\hat{\bm{\sigma}}$ are assumed to be the block matrices with their components being matrices acting on the mass states.

The matrix element of the transition $\nu_i+N\to\nu_n+N$ due to the weak interaction is given by
\begin{eqnarray}
\label{M_weak}
     \mathcal{M}^{(w)}&=-\dfrac{G_F}{\sqrt{2}}\bar{u}^{(\nu_n)}_{k',r'}\gamma^\mu(1-\gamma^5)\delta^{ni}u^{(\nu_i)}_{k,r}J^{(NC)}_\mu(q),
\end{eqnarray}
where $J_\lambda^{(NC)}(q)$ is the neutral weak nucleon current.
The matrix element due to the EM interaction is given by
\begin{eqnarray}
\label{M_el-m}
     \mathcal{M}^{(\gamma)}&=\dfrac{4\pi\alpha}{q^2}\bar{u}^{(\nu_n)}_{k',r'}\Lambda^{({\rm EM};\nu)ni}_\mu(q)u^{(\nu_i)}_{k,r}J^{(EM)\mu}(q),
\end{eqnarray}
where $J_\mu^{(EM)}(q)$ is the EM nucleon current. Assuming the target nucleon to be free, these nucleon transition currents can be parameterized as follows:
\begin{equation}
\label{currents1}
\begin{aligned}
     J^{(NC)}_\mu(q)&=\bar{u}^{(N)}_{p',s'}\Lambda^{({\rm NC};N)}_\mu(-q)u^{(N)}_{p,s}\\
     &=F_1^N(q)J^V_\mu(q)-G_A^N(q)J^A_\mu(q)+\frac{i}{2m_N}F_2^N(q)J^M_\mu(q),\\
     J^{(EM)}_\mu(q)&=\bar{u}^{(N)}_{p',s'}\Lambda^{({\rm EM};N)}_\mu(-q)
     u^{(N)}_{p,s}\\
     &=F_Q(q)^NJ^V_\mu(q)-\left(\delta_\mu^\lambda q^2-q^\lambda q_\mu\right)\frac{F_A^N(q)}{m_N^2}J^A_\lambda(q)+\frac{i}{2m_N}F_M^N(q)J^M_\mu(q)-\frac{F_E^N(q)}{2m_N}J^E_\mu(q),
\end{aligned}
\end{equation}
where
\begin{equation}
\label{J_V_A M E}
\begin{aligned}
     J^V_\mu(q)&=\bar{u}^{(N)}_{p',s'}\gamma_\mu u^{(N)}_{p,s}, \qquad
     J^A_\mu(q)=\bar{u}^{(N)}_{p',s'}\gamma_\mu\gamma_5 u^{(N)}_{p,s}, \\
     J^M_\mu(q)&=\bar{u}^{(N)}_{p',s'}\sigma_{\mu\alpha}q^\alpha u^{(N)}_{p,s} \qquad
     J^E_\mu(q)=\bar{u}^{(N)}_{p',s'}\sigma_{\mu\alpha}q^\alpha\gamma_5 u^{(N)}_{p,s},
\end{aligned}
\end{equation}
with $u^{(N)}_{p,s}$ being the bispinor amplitude of the nucleon with four-momentum $p$ and spin $s$.

Since we are interested in the scattering of ultrarelativistic neutrinos, we disregard the neutrino mass in our calculations of the cross section. In the lab frame with $z$-axis directed along the incident neutrino momentum we calculate the angular differential cross section
\begin{equation}
\label{cr_sec}
     \frac{d\sigma}{d\Omega}=\frac{d^2\sigma}{\sin{\theta}d\theta d\varphi}=\frac{\left|\mathcal{M}\right|^2t^2}{64\pi^2(s-m^2_N)^2m^2_N}\left(1-\frac{4m^2_N}{t}\right)^{3/2}
\end{equation}
in terms of the Mandelstam variables 
$$
s=(k+p)^2=m^2_N+2E_\nu m_N, \qquad t=(k-k')^2=-\frac{4m^2_NE^2_\nu\cos^2{\theta}}{(E_\nu+m_N)^2-E^2_\nu\cos^2{\theta}}
$$ 
as a function of the polar ($\theta$) and azimuthal ($\varphi$) angles of the recoil nucleon. Also, averaging over initial and summing over final spin polarizations of the nucleon ($\frac{1}{2}\sum_{s's}$) and summing over final neutrino polarisations ($\sum_{r'}$) and final neutrino mass states ($\sum_f$) are assumed.
The absolute matrix element squared for the neutrino initial and final states $\rho_{ij}$ and $\rho^{f}_{n'n}$ has the following structure:
%
\begin{equation}
\label{M01}
\begin{aligned}
     \left|\mathcal{M}\right|^2&=\frac{G_F^2}{2}\sum_{i,j,n,n'=1}^3\frac{1}{2}\sum_{s's}\sum_{r'}\sum_f tr\{\rho^{f}_{n'n}\mathcal{O}^{s's}_{ni}\rho_{ij}\gamma^0(\mathcal{O}^{s's}_{n'j})^\dagger\gamma^0\},
\end{aligned}
\end{equation}
%
where $\mathcal{O}^{s's}_{ni}$ describes interactions with the nucleon neutral weak and EM currents
\begin{equation}
    \mathcal{O}^{s's}_{ni}=\delta_{ni}\gamma^\mu(1-\gamma_5)J^{s's({\rm NC})}_\mu(q)-\frac{4\sqrt{2}\pi\alpha}{G_F^2q^2}\Lambda^{({\rm EM};\nu)\mu}_{ni}(q)J^{s's(EM)}_\mu(q).
\end{equation}
In the calculation, we take into account that $\sum_{r'}\sum_{f}\rho^{f}_{n'n}=\delta_{n'n}\slashed{k}'$.
Therefore, the cross section can be split into the following parts:
\begin{equation}
\label{Full differential cross section}
     \frac{d\sigma}{d\Omega}=\frac{d\sigma^L}{d\Omega}+\frac{d\sigma^R}{d\Omega}+\frac{d\sigma^\perp}{d\Omega},
\end{equation}
where $\frac{d\sigma^K}{d\Omega}$ are cross sections with projection of the initial neutrino state on the left ($K=L$) and right ($K=R$) helicity states, respectively. Both of them can be divided into helicity-preserving and helicity-flipping components (to avoid encumbering the presentation, we omit the argument $q^2=t$ in the expressions for the form factors); that is,
%
\begin{equation}
\label{cross sectionLR}
\begin{aligned}
     \frac{d\sigma^K}{d\Omega}=&\frac{d\sigma^K_{\rm hp}}{d\Omega}+\frac{d\sigma^K_{\rm hf}}{d\Omega},\\
     \frac{d\sigma^K_{\rm hp}}{d\Omega}=&\frac{G^2_Ft^2(s-m^2_N)}{16\pi^2m^2_N(s+m^2_N)}\left(1-\frac{4m^2_N}{t}\right)^{3/2}\\
     &\times\Bigg[2\left(1+\frac{st}{(s-m^2_N)^2}\right)\left(C^K_V+C^K_A-\frac{t}{4m^2_N}(C^K_M+C^K_E)\right)\\
     &-\frac{4m^2_Nt}{(s-m^2_N)^2}\left(C^K_A-\frac{t}{4m^2_N}C^K_M\right)+\frac{t^2}{(s-m^2_N)^2}\left(C^K_V+C^K_A-2{\rm Re}\,C^K_{V\&M}\right)\\
     &\pm\frac{2t}{s-m^2_N}\left(2+\frac{t}{s-m^2_N}\right){\rm Re}\,(C^K_{V\&A}-C^K_{A\&M})\Bigg],\\
     \frac{d\sigma^K_{\rm hf}}{d\Omega}=&\frac{\alpha^2t^2(s-m^2_N)}{8m^2_em^3_N(s+m^2_N)}\left(1-\frac{4m^2_N}{t}\right)^{3/2}|\mu^K_\nu|^2\Bigg[-\frac{2m_N}{t}\left(1+\frac{t}{s-m^2_N}\right)(F^N_Q)^2\\
     &-\frac{2t}{m^3_N}\left(1+\frac{st}{(s-m^2_N)^2}\right)(F^N_A)^2+\frac{m_Nt}{(s-m^2_N)^2}F^N_QF^N_M\\
     &+\frac{1}{8m_N}\left(4+\frac{4st+t^2}{(s-m^2_N)^2}\right)(F^N_M)^2+\frac{1}{8m_N}\left(2+\frac{t}{s-m^2_N}\right)^2(F^N_E)^2\Bigg],
\end{aligned}
\end{equation}
%
where $+$ ($-$) stands for $K=L$ ($K=R$) and~\cite{nue}
%
\begin{equation}
\label{C-coefficients}
\begin{aligned}
     &C^K_V=Tr\left[\left(-F_1^N\delta^K_L+F_Q^NQ^K\right)^2\rho^K\right], \qquad C^K_A=Tr\left[\left(\delta^K_LG^N_A-\dfrac{tF^N_AQ^K}{m_N^2}\right)^2\rho^K\right],\\
     &C^K_{V\&A}=Tr\left[\left(\delta^K_LG^N_A-\dfrac{tF^N_AQ^K}{m_N^2}\right)\left(-F_1^N\delta^K_L+F_Q^NQ^K\right)\rho^K\right], \quad C^K_M=Tr\left[\left(\delta^K_LF^N_2-F^N_MQ^K\right)^2\rho^K\right],\\
     &C^K_{V\&M}=Tr\left[\left(\delta^K_LF^N_2-F^N_MQ^K\right)\left(-F_1^N\delta^K_L+F_Q^NQ^K\right)\rho^K\right],
     \qquad  C^K_E=Tr\left[\left(F_EQ^K\right)^2\rho^K\right],\\
     &C^K_{A\&M}=Tr\left[\left(\delta^K_LF^N_2-F^N_MQ^K\right)\left(\delta^K_LG^N_A-\dfrac{tF^N_AQ^K}{m_N^2}\right)\rho^K\right],\\    
     &\rho^{L,R}=\frac{1}{2}(\tilde\rho\mp\zeta^\parallel), \quad Q^{L,R}=\frac{2\sqrt{2}\pi\alpha}{G_Ft}\left(f^Q\mp tf^A\right), \quad  |\mu^{L,R}_\nu|^2=Tr\left[\left(f^M\pm if^E\right)\left(f^M\mp if^E\right)\rho^{L,R}\right].
\end{aligned}
\end{equation}
%
The cross section $\frac{d\sigma^\perp}{d\Omega}$ describes the contribution from the interference of the helicity states and depends on the transverse neutrino spin components $\zeta^x_{ij}$ and $\zeta^y_{ij}$ written in terms of $\kappa_{ij}=\frac{1}{2}(\zeta^x_{ij}+i\zeta^y_{ij})$ [see the text after Eq.~(\ref{Landau density matrix})]
%
\begin{equation}
\label{perp cross section}
\begin{aligned}
    \frac{d\sigma^\perp}{d\Omega}=&-\frac{\sqrt{2}G_F\alpha(s-m^2_N)\left(4m^2_N-t\right)^{3/2}}{8\pi m_em^2_N(s+m^2_N)}\sqrt{1+\frac{st}{(s-m^2_N)^2}}\Bigg\{\frac{2t}{s-m^2_N}\frac{tF^N_A}{m^2_N}(F^N_Q+F^N_M)C^+_{\mu_\nu,Q}\\
    &+C^\perp_{\mu_\nu,Z^0}\left[\left(2+\frac{t}{s-m^2_N}\right)\left(F^N_1F^N_Q+\frac{tF^N_A}{m^2_N}G^N_A-t\frac{F^N_2F^N_M}{4m^2_N}\right)\right.\\
    &\left.-\frac{t}{s-m^2_N}\left(\frac{tF^N_A}{m^2_N}(F^N_1+F^N_2)+G^N_A(F^N_Q+F^N_M)\right)\right]\\
     &-\left(2+\frac{t}{s-m^2_N}\right)\left(\left(F^N_Q\right)^2+\left(\frac{tF^N_A}{m^2_N}\right)^2-t\left(\frac{F^N_M}{2m_N}\right)^2-t\left(\frac{F^N_E}{2m_N}\right)^2\right)C^-_{\mu_\nu,Q}\Bigg\},
\end{aligned}
\end{equation}
%
where
\small
\begin{equation}
\label{Perp-coefficients}
\begin{aligned}
     &C^\perp_{\mu_\nu,Z^0}={\rm Re}\,Tr\left[\left(f^M+if^E\right)\kappa^\dagger e^{i\varphi}\right], \\
     &C^\pm_{\mu_\nu,Q}={\rm Re}\,Tr\left[\left[Q^L\left(f^M+if^E\right)\pm\left(f^M+if^E\right)Q^R\right]\kappa^\dagger e^{i\varphi}\right].
\end{aligned}
\end{equation}
\normalsize
It is important to note that the cross sections $\frac{d\sigma^{L,R}}{d\Omega}$ depend only on the polar angle $\theta$ and are independent of the transverse neutrino polarization components. Also, only the EM interaction contributes to the cross section for right-handed neutrinos ($K=R$). The cross section $\frac{d\sigma^\perp}{d\Omega}$ depends both on the polar and azimuthal angles of the recoil nucleon and involves the terms due to interference between the weak and EM interactions.

\section{Numerical results}
\label{results}
Obviously, manifestations of neutrino EM properties should be much more pronounced in neutrino-proton rather than neutrino-neutron scattering. Indeed, our calculations show that the relative contribution of the neutrino magnetic moment (charge radius or anapole moment) to the neutrino-nucleon cross section in the case of a neutron is about eight (three) orders of magnitude smaller than in the case of a proton. Therefore, in order to illustrate the characteristic effects of these properties, we present below the results obtained by numerically calculating the differential cross section for elastic neutrino-proton scattering.

The calculations have been made not only for the cross section $\frac{d\sigma}{d\Omega}=\frac{d^2\sigma}{\sin{\theta}d\theta d\varphi}$ given by Eqs.~(\ref{Full differential cross section})-(\ref{Perp-coefficients}), but also for the cross section $\frac{d\sigma}{dT}$, which is differential with respect to the energy transfer $T$ and which determines the shape of the nucleon recoil spectrum. The cross sections are related to each other in accordance with
\begin{equation}
\label{ThetaTRelatio}
    \begin{aligned}
        \cos\theta=\sqrt{\frac{T}{T+2m_N}}\frac{E_\nu+m_N}{E_\nu}, \qquad \frac{d\sigma}{dT}=\frac{m_N(E_\nu+m_N)}{E_\nu\sqrt{T}(T+2m_N)^{3/2}}\int_0^{2\pi}\frac{d\sigma}{d\Omega}d\varphi.
    \end{aligned}
\end{equation}
\subsection{Neutrino charge radii}
First, we present numerical calculations for cross sections differential with respect to the energy transfer $T$ with and without allowance for the neutrino charge radii. The calculations are performed for neutrino energies typical for supernova sources. It is generally believed that, due to the neutrino flavor oscillations and their decoherence on a large source-detector base, the supernova neutrino should arrive at the detector as an incoherent mixture of three flavor states ($\ell=e,\mu,\tau$). For each neutrino flavor three different spin states are considered: the arriving neutrino can be either left-handed, right-handed, or fully unpolarized. The SM results are obtained both with and without taking into account the strange contribution to the proton form factors, which was treated according to the approach outlined in Sec.~\ref{Parametrization of nucleon form factors}. In the case of the diagonal neutrino charge radii, we used the SM values~(\ref{SM CR values}) and~(\ref{SM vAnapole}), taking into account the arbitrariness of the interpretation of the neutrino charge radius and anapole moment in the SM. However, their effect is so small that the corresponding results are visually indistinguishable from the results without them. Therefore, fig.~\ref{TCR} shows the effect of the neutrino transition charge radii. For the latter we employed the values (in cm$^2$) in the range which is consistent with the current experimental constraints~\cite{Cadeddu2020,Dresden-II,PDG24}:
\begin{equation}
\label{CR_transit}
    0\leq|\langle r_{e\mu}^2\rangle|,|\langle r_{e\tau}^2\rangle|,|\langle r_{\mu\tau}^2\rangle|<3\times10^{-31}.
\end{equation}
In such a case the results do not depend on the neutrino flavor.
\begin{center}
	   {\includegraphics[width=0.45\linewidth]{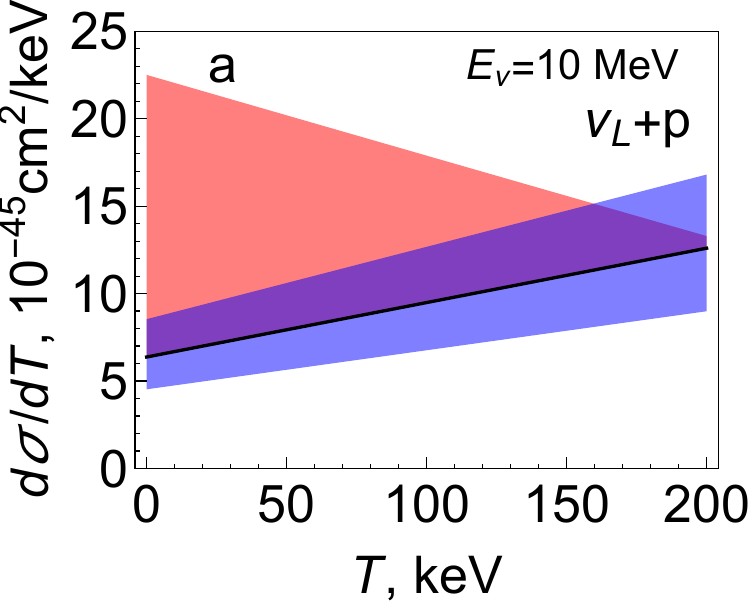}}
	\hfill
	   {\includegraphics[width=0.45\linewidth]{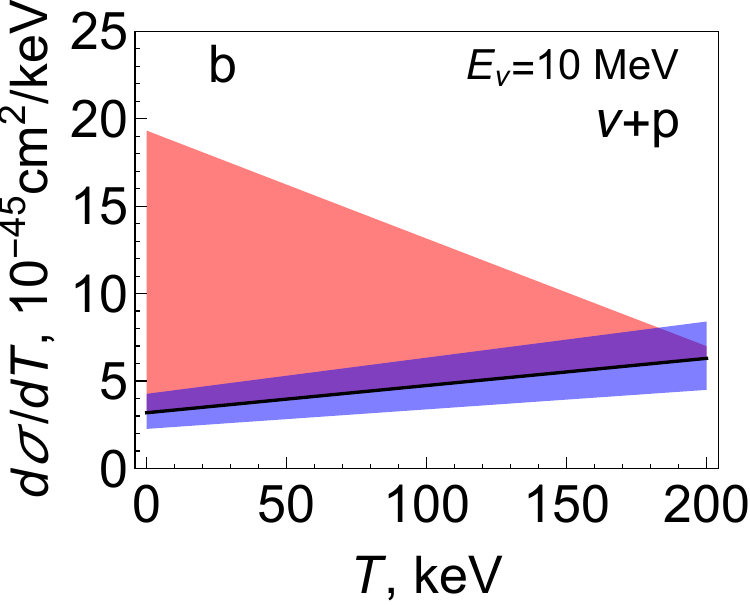}}
    \hfill
	   \center{\includegraphics[width=0.45\linewidth]{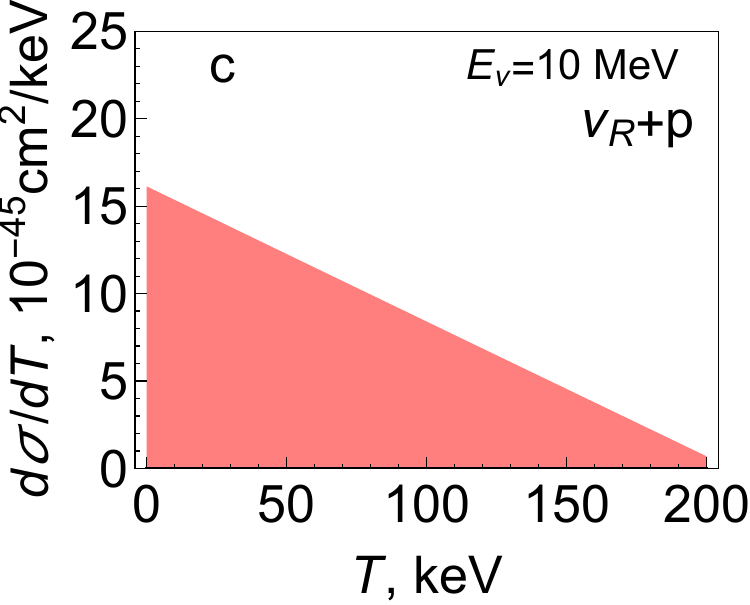}}
    \center{\includegraphics[width=0.65\linewidth]{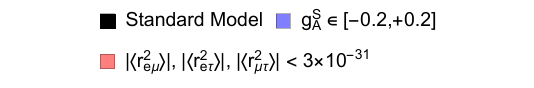}}
	\captionof{figure}{The effect of the neutrino transition charge radii on the differential cross section for elastic neutrino-proton scattering in the cases of different initial neutrino spin states in the detector: (a) left-handed, (b) fully unpolarazed and (c) right-handed. The neutrino energy is $E_\nu=10$ MeV, which is typical for a supernova source.}
	\label{TCR}
\end{center}

\subsection{Neutrino anapole moments}
We also examine possible contributions from neutrino anapole moments. Similary to fig.~\ref{TCR}, the SM predictions for diagonal anapole moments~(\ref{SM vAnapole}) bring about only minor corrections, which cannot be visually distinguished from the results without them. On the other hand, the current neutrino scattering experiments limit the effective charge radii $\langle r_{ij}^2\rangle_L=\langle r_{ij}^2\rangle-6a_{ij}$, which are linear combinations of the neutrino charge radii and anapole moments. Namely, one shoud make the substitution $\langle r^2_{ij}\rangle\rightarrow\langle r_{ij}^2\rangle_L$ in Eq.~(\ref{CR_transit}). Therefore, when we calculate the effect of transition anapole moments while assuming zero charge radii, we obtain the results identical to those in fig.~\ref{TCR}. 

Figure~\ref{TCRAM} shows the calculations with the neutrino charge radii and anapole moments, provided the SM-like relationship~(\ref{SM vAnapole}) between them holds, i.e.,
\begin{equation}
\label{BSM vAnapole}
    a_{ij}=-\frac{1}{6}\langle r^2_{ij}\rangle,\qquad i,j=\{1,2,3\}.
\end{equation}
\begin{center}
	   \includegraphics[width=0.45\linewidth]{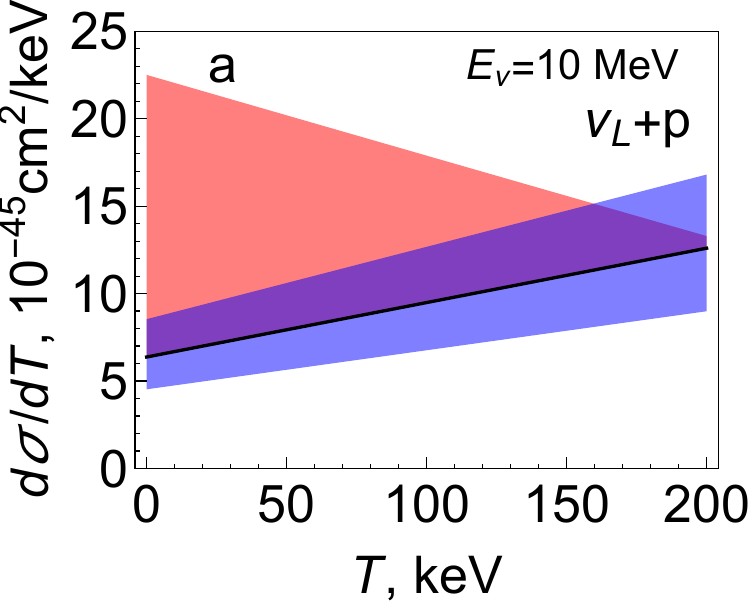}
	\hfill
	\includegraphics[width=0.45\linewidth]{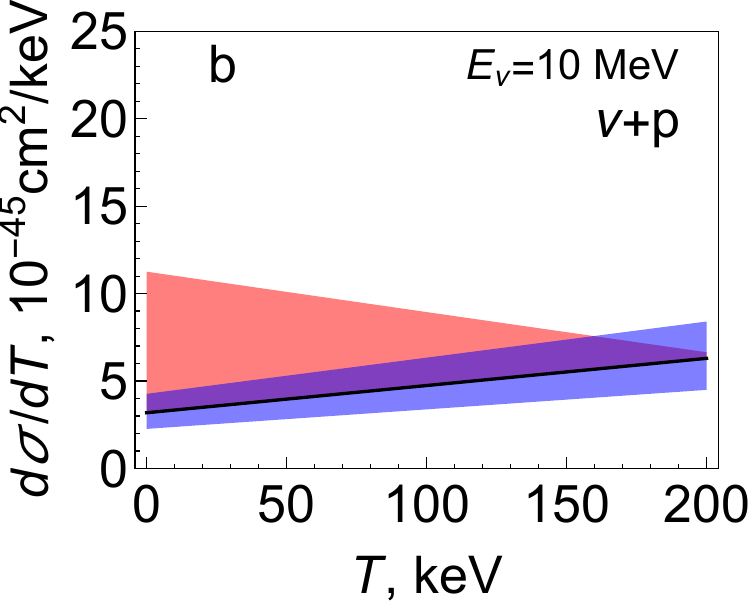}
    \hfill
	   \center{\includegraphics[width=0.45\linewidth]{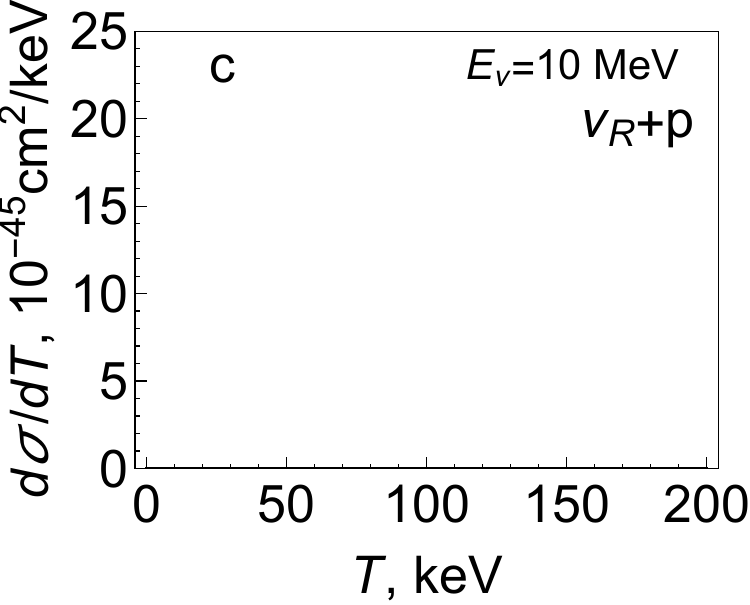}}
    \center{\includegraphics[width=0.65\linewidth]{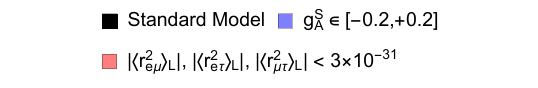}}
	\captionof{figure}{The same as in fig.~\ref{TCR}, but in the case of the effective neutrino transition charge radii.}
	\label{TCRAM}
\end{center}
Figure~\ref{TCR} demonstrates that neutrino EM properties reveal themselves considerably for both left- and right-handed neutrinos, whereas fig.~\ref{TCRAM} demonstrates that the effects of charge radii and anapole moments can be distinguished. Thus, due to Eq.~(\ref{BSM vAnapole}) the cross section for left-handed neutrinos depends on the linear combination $\langle r_{ij}^2\rangle_L=\langle r_{ij}^2\rangle-6a_{ij}=2\langle r_{ij}^2\rangle$ and exhibits the maximum EM effect, while for right-handed neutrinos the cross section depends on the linear combination $\langle r_{ij}^2\rangle_R=\langle r_{ij}^2\rangle+6a_{ij}=0$ and there is no EM effect at all. At the same time, it should be noted that in the general case of beyond-SM theories the relation~(\ref{BSM vAnapole}) does not necessarily hold and the neutrino charge radius and anapole moment can arise as independent EM characteristics.

\subsection{Neutrino magnetic moments}
Figure~\ref{MMl} shows the effects of the neutrino magnetic moments (in units of $\mu_B$) lying in the ranges \cite{LUX23}
\begin{equation}
\label{mu_nu_diag}
    |\mu_{ee}|<1.5\times10^{-11}, \qquad |\mu_{\mu\mu}|<2.3\times10^{-11}, \qquad |\mu_{\tau\tau}|<2.1\times10^{-11},
\end{equation}
\begin{center}
    \begin{minipage}[h]{0.32\linewidth}
	   \center{\includegraphics[width=\linewidth]{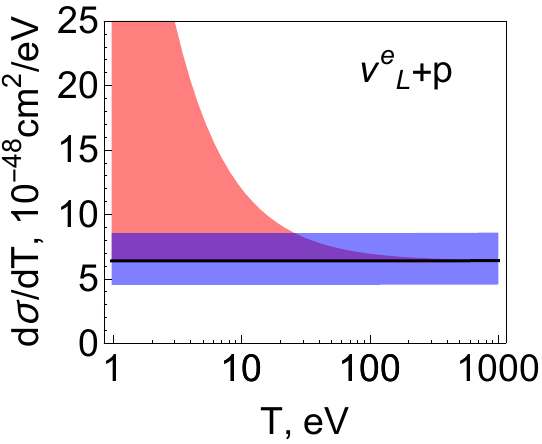}}
	\end{minipage}
	\begin{minipage}[h]{0.32\linewidth}
	   \center{\includegraphics[width=\linewidth]{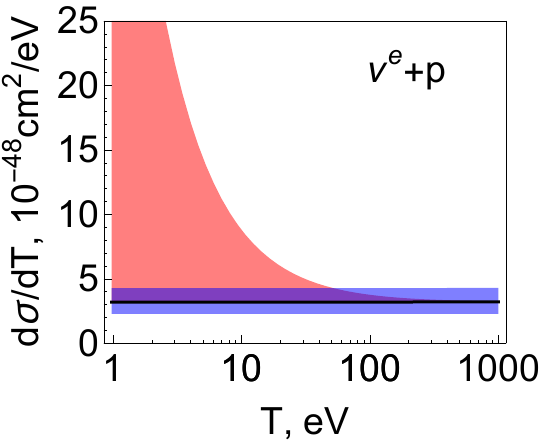}}
	\end{minipage}
	\begin{minipage}[h]{0.32\linewidth}
        \center{\includegraphics[width=\linewidth]{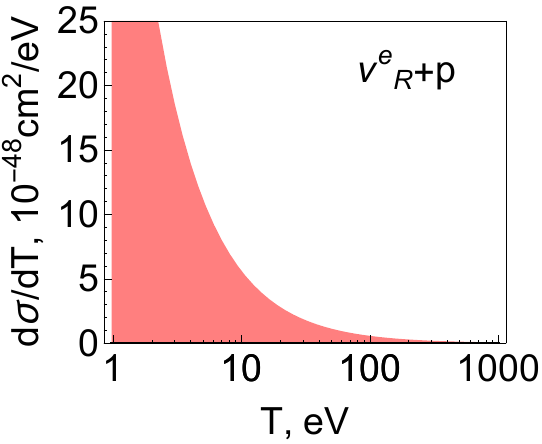}}
	\end{minipage}
    \vspace{0.5cm}
    \begin{minipage}[h]{0.32\linewidth}
        \center{\includegraphics[width=\linewidth]{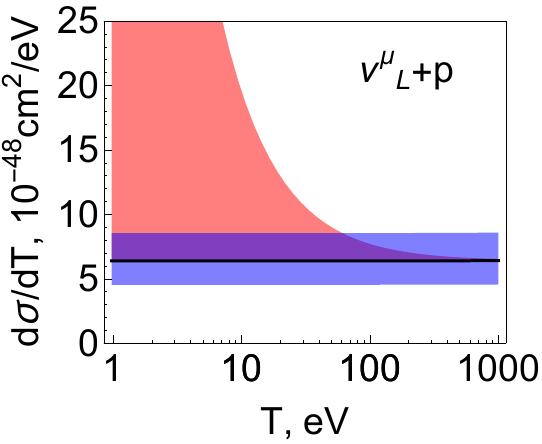}}
	\end{minipage}
	\begin{minipage}[h]{0.32\linewidth}
        \center{\includegraphics[width=\linewidth]{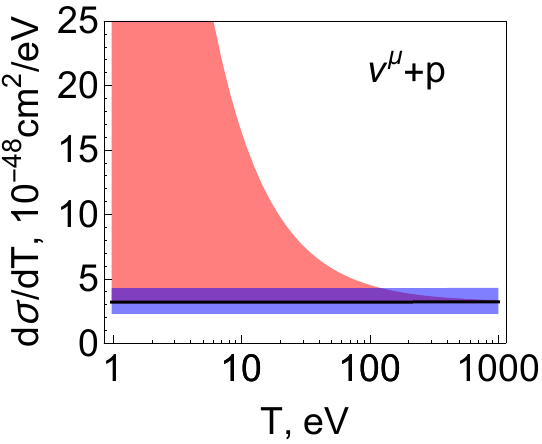}}
	\end{minipage}
	\begin{minipage}[h]{0.32\linewidth}
        \center{\includegraphics[width=\linewidth]{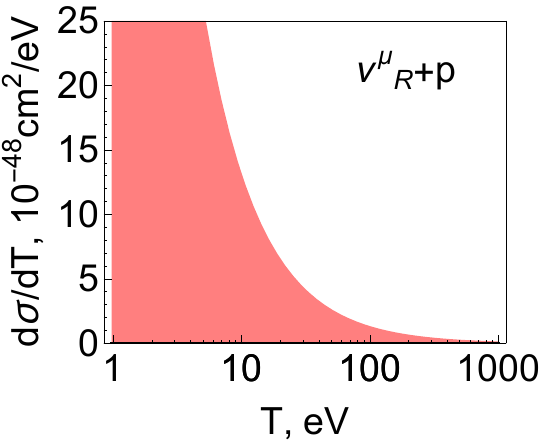}}
	\end{minipage}
    \vspace{0.5cm}
    \begin{minipage}[h]{0.32\linewidth}
        \center{\includegraphics[width=\linewidth]{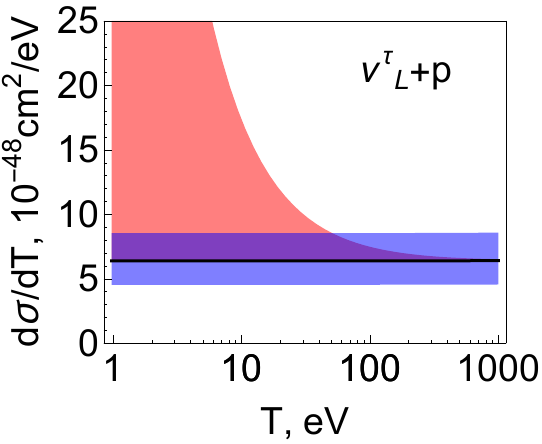}}
	\end{minipage}
	\begin{minipage}[h]{0.32\linewidth}
        \center{\includegraphics[width=\linewidth]{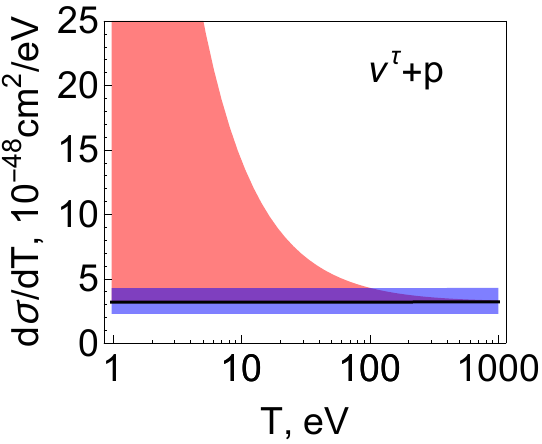}}
	\end{minipage}
	\begin{minipage}[h]{0.32\linewidth}
        \center{\includegraphics[width=\linewidth]{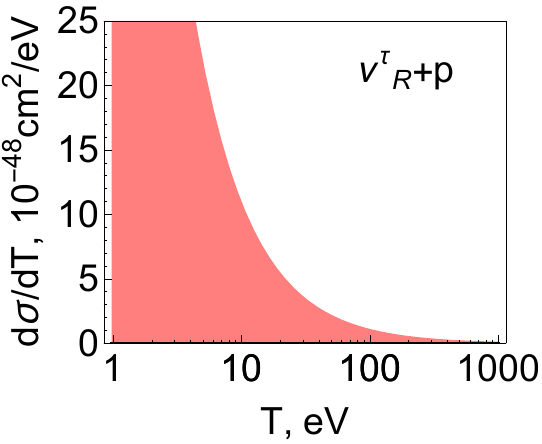}}
	\end{minipage}
    \center{\includegraphics[width=0.65\linewidth]{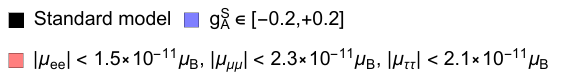}}
	\captionof{figure}{The differential cross section for elastic neutrino-proton scattering in the cases of different initial neutrino states in the detector. Top, middle and bottom rows correspond to the electron, muon and tau flavors, respectively. Left, middle and right columns correspond to left-handed, fully unpolarizaed and right-handed neutrinos, respectively.}
	\label{MMl}
\end{center}
which also reflect typical values of upper limits obtained for the neutrino magnetic moment from various neutrino scattering experiments~\cite{Cadeddu2020,Dresden-II,PDG24,CONUS2022,Xenon1t}. 
Since for a nonzero neutrino magnetic moment the cross section $\frac{d\sigma}{dT}$ is singular when $T\to0$, we show the region of low energy transfers, $T\ll E_\nu$. It should be noted that in such kinematical regime the cross section is practically independent of $E_\nu$. 

\subsection{Neutrino transverse spin polarization}
Figure~\ref{MMt} ilustatrates the effect of the neutrino transverse spin polarization on the cross section differential with respect to the solid angle of the recoil proton. Let us note that only when there is a transverse component of the neutrino spin polarization the cross section depends on the azimuthal angle of the recoil neucleon. We assume the initial neutrino to be fully transversely polarized relative to the initial neutrino momentum 
and compare this case with that when the neutrino is fully unpolarized. As follows from Eq.~(\ref{perp cross section}), the contribution from the neutrino transverse spin polarization to the cross section is proportional to the neutrino magnetic and/or electric dipole moments, so that it completely vanishes if they have zero values. The effect turns out to be on such a tiny scale that the contributions from the neutrino charge radii and anapole moments within the SM also become visible, and, moreover, they tend to amplify the discussed effect. 

\begin{center}
    \begin{minipage}[h]{0.45\linewidth}
	   \center{\includegraphics[width=\linewidth]{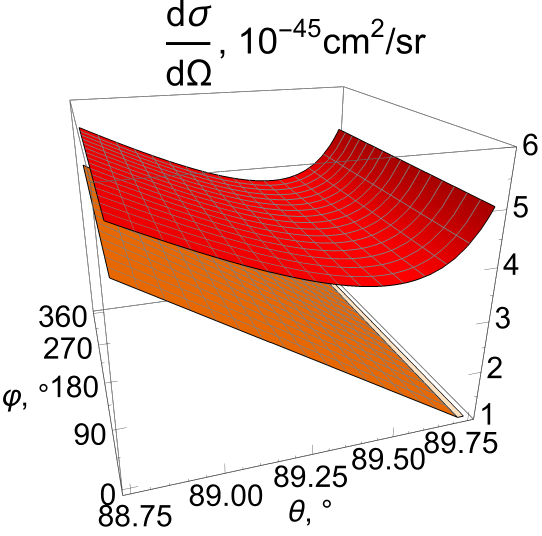}}
    \end{minipage}
    \begin{minipage}[h]{0.45\linewidth}
	   \center{\includegraphics[width=\linewidth]{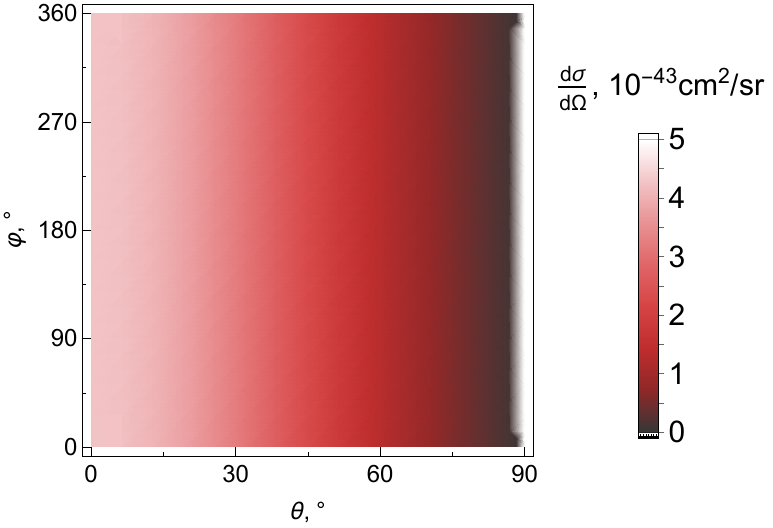}}
    \end{minipage}\\
    \vspace{0.5cm}
    \begin{minipage}[h]{0.45\linewidth}
	   \center{\includegraphics[width=\linewidth]{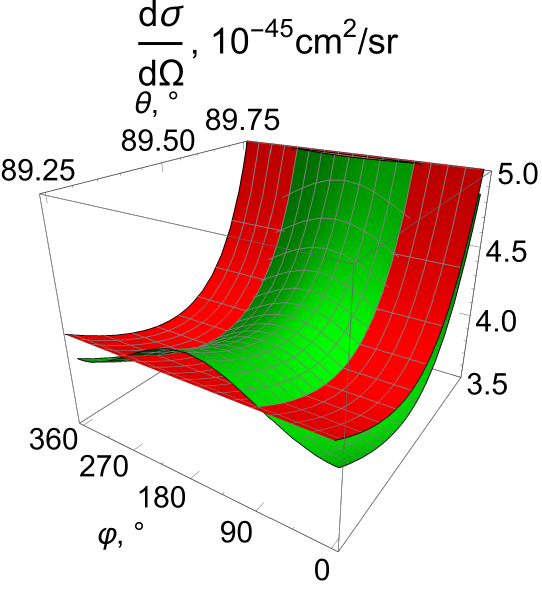}}
    \end{minipage}
    \begin{minipage}[h]{0.45\linewidth}
	   \center{\includegraphics[width=\linewidth]{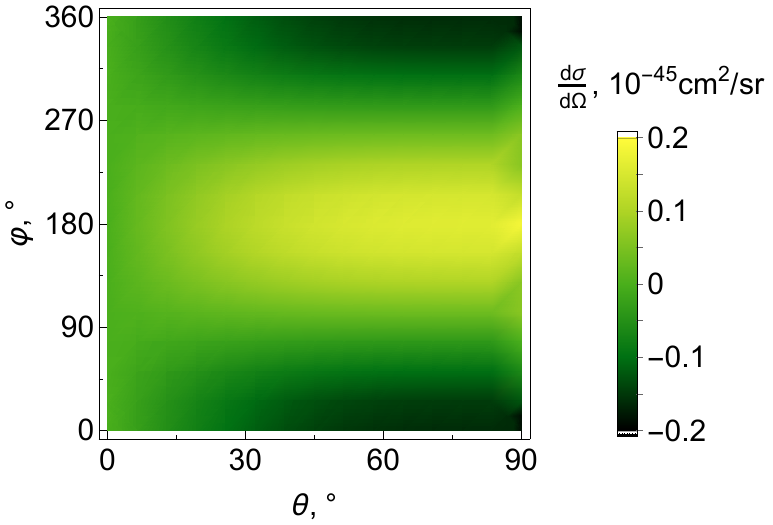}}
    \end{minipage}\\
    \vspace{0.5cm}
    \begin{minipage}[h]{0.45\linewidth}
	   \center{\includegraphics[width=\linewidth]{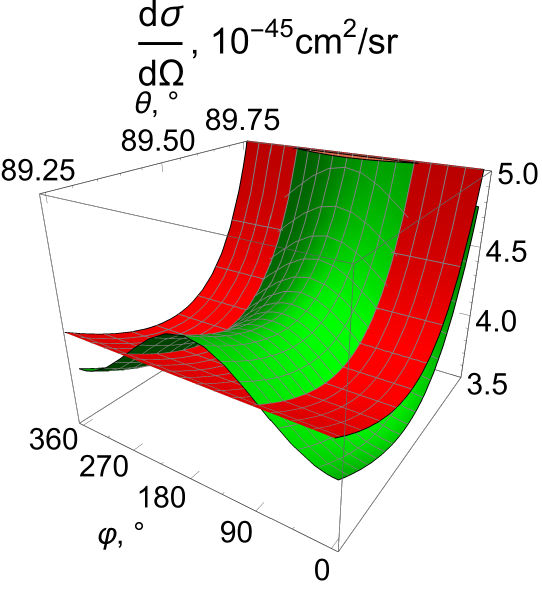}}
    \end{minipage}
    \begin{minipage}[h]{0.45\linewidth}
	   \center{\includegraphics[width=\linewidth]{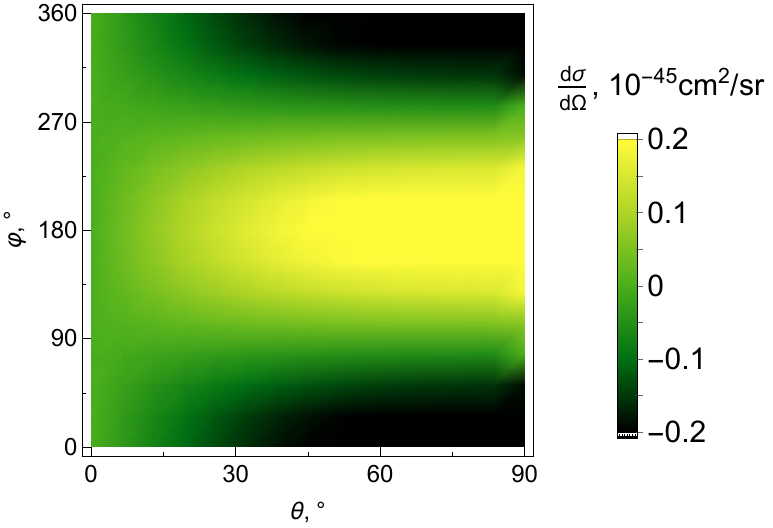}}
    \end{minipage}\\
    \center{\includegraphics[width=0.65\linewidth]{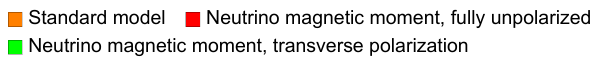}}
	\captionof{figure}{The angular differential cross section for elastic neutrino-proton scattering in the cases of different initial neutrino spin states in the detector. In the first row, the 3d plot presents cross sections for fully unpolarized neutrinos (i) within the SM and (ii) accounting for neutrino magnetic moments of $10^{-11}\mu_B$, and the 2d density plot shows only the latter cross section. In the second row, the 3d plot compares the cross sections accounting for neutrino magnetic moments of $10^{-11}\mu_B$ in the cases of fully unpolarized and transversely polarized neutrinos, while the 2d density plot shows the difference between these cross sections. Finally, the third row is the same as the second row, but taking into account the effect of the SM charge radii and anapole moments.}\label{MMt}
\end{center}

\section{Summary and conclusions}
\label{summary}
The theory of elastic neutrino-nucleon scattering has been developed taking into account the EM interactions of massive neutrinos with arbitrary spin polarizations. The process under study proceeds via two interaction channels, namely via the $Z^0$-boson and photon exchange. In either case, the nucleon form factors have been taken into account within the developed formalism. In addition, we have taken into account the mixing, oscillations and polarization of neutrinos propagating from the source to the target. We have obtained general expressions for the differential cross section for elastic neutrino-nucleon scattering and have employed them to perform numerical calculations for the scattering of supernova neutrinos on a proton target. Also, we have outlined possible manifestations of the effects of the neutrino magnetic moments, transition (in the flavor basis) charge radii and anapole moments on the cross section for left- and right-handed initial neutrino states, as well as the effects of the transverse neutrino spin polarization.

From the obtained numerical results, one can see that the neutrino EM properties can make a substantial contribution to the cross section for elastic neutrino-proton scattering. At the same time, an implementation of full-scale searches for these properties would require experimentally measuring the differential cross section over a rather wide energy-transfer range. This would permit (i) separating the effects of neutrino EM interactions from the effects associated with nucleon form factors and (ii) pinpointing neutrino EM properties that are responsible for the observed effects.

Also, our analytical and numerical calculations clearly demonstrate that the neutrino EM properties can manifest themselves both for left- and right-handed neutrinos. This means, in particular, that the fluxes of right-handed neutrinos coming to the detector should also be taken into account. In order to point out this feature, in our numerical calculations we have 
considered nine different incoming neutrino states: 3 flavor states $\times$ 3 spin states (left-handed, right-handed and fully unpolarized). 
In the general case, the initial neutrino state in the detector should be described by the density matrix in the space of a tensor product of flavor (or mass) and spin subspaces. The density matrix is determined by the neutrino propagation from the source to the detector, which depends on the neutrino interactions with matter and EM fields during the propagation. In such a general case, we have shown that even the transverse neutrino spin component can manifest itself in the neutrino-nucleon scattering process. On the other hand, the corresponding effect appears to be rather small. 
Given the structure of the derived angular differential cross section, we might expect a much larger influence of the transverse neutrino spin component in the case of nuclear targets that are much heavier than a nucleon.

Finally, the expressions for the cross sections obtained in this work carry information about both the neutrino and nucleon EM form factors. This makes it possible to apply these expressions in analyzing the results of various experiments aimed at studying neutrino interactions and oscillations in matter, detecting neutrinos from supernova explosions on the basis of elastic neutrino-proton scattering, measuring the nucleon anapole moment, and searching for the neutron electric dipole moment. In addition, the results of the present study contribute to the development of a systematic approach to studying the EM properties of neutrinos in their scattering on composite targets (nuclei, atoms, and condensed media).

\begin{acknowledgments}
This work is supported by the Russian Science Foundation (project No. 24-12-00084).
\end{acknowledgments}

\bibliographystyle{apsrev}
\bibliography{ref}
\end{document}